\documentclass[smallabstract,smallcaptions]{dccpaper}

\usepackage{epsfig}
\usepackage{citesort}
\usepackage{amsmath}
\usepackage{amssymb}
\usepackage{color}
\usepackage{url}

\usepackage{soul}
\usepackage{cite}

\newlength{\figurewidth}
\newlength{\smallfigurewidth}

\usepackage{caption}
\usepackage{subcaption}
\usepackage{hyperref}
\usepackage{booktabs}
\usepackage{array,multirow}

\def\Resize{\texttt{Resize}}

\DeclareMathOperator{\Redundancy}{Rdn}

\def\RedundancyMetric#1#2{%
  \Redundancy{}(%
    \boldsymbol{y}_{#1} \mid
    \boldsymbol{y}_{#2}%
  )%
}

\def\EstimatedRedundancyMetric#1#2{%
  \widetilde{\Redundancy{}}(%
    \boldsymbol{y}_{#1} \mid
    \boldsymbol{y}_{#2}%
  )%
}

\def\MIRatio#1#2{%
  \frac{%
    I(\boldsymbol{y}_{#1} ; \boldsymbol{y}_{#2})%
  }{%
    H(\boldsymbol{y}_{#1})%
  }%
}

\def\FullEntropyRatio#1#2{%
  \frac%
  {H(\boldsymbol{y}_{#1} \mid \boldsymbol{y}_{#2})}%
  {H(\boldsymbol{y}_{#1})}%
}

\def\EntropyRatio#1#2{\textrm{ER}_{#1#2}}

\def\EntropyRatio#1#2{\FullEntropyRatio{#1}{#2}}

\DeclareMathAlphabet\mathbfcal{OMS}{cmsy}{b}{n}

\begin{document}

\title
{\large
\textbf{Learned Disentangled Latent Representations for Scalable Image Coding for Humans and Machines}
}

\author{%
Ezgi {\"O}zyılkan$^{1, {\dag, \ast}}$, Mateen Ulhaq$^{1, {\ddag, \ast}}$, Hyomin Choi$^{\ast}$, and Fabien Racapé$^{\ast}$\\
{\small\begin{minipage}{\linewidth}\begin{center}
\begin{tabular}{c}
$^{\dag}$Dept.~of Electrical and Computer Engineering, New York University \\
$^{\ddag}$School of Engineering Science, Simon Fraser University \\
$^{\ast}$ InterDigital - Emerging Technologies Lab \\
eo2135@nyu.edu, mulhaq@sfu.ca, \{hyomin.choi, fabien.racape\}@interdigital.com
\end{tabular}
\end{center}\end{minipage}}
}

\maketitle
\textbf{\footnotetext[1]{Contributed equally to this work. \\This work was  done while E. {\"O}zyılkan and M. Ulhaq were interns at InterDigital.}}
\thispagestyle{empty}

\begin{abstract}
As an increasing amount of image and video content will be analyzed by machines, there is demand for a new codec paradigm that is capable of compressing visual input primarily for the purpose of computer vision inference, while secondarily supporting input reconstruction.
In this work, we propose a learned compression architecture that can be used to build such a codec.
We introduce a novel variational formulation that explicitly takes feature data relevant to the desired inference task as input at the encoder side.
As such, our learned scalable image codec encodes and transmits two disentangled latent representations for object detection and input reconstruction.
We note that compared to relevant benchmarks, our proposed scheme yields a more compact latent representation that is specialized for the inference task. Our experiments show that our proposed system achieves a bit rate savings of 40.6\% on the primary object detection task compared to the
current state-of-the-art, albeit with some degradation in performance for the secondary input reconstruction task.

\end{abstract}

\section{Introduction}

It is projected that an increasing amount of captured visual content will be analyzed by \emph{machines} in order to conduct vision analytics (e.g., object detection, image classification, segmentation), instead of being solely viewed by humans~\cite{CISCO_VNI_2023}.
Since recent advances in artificial intelligence using deep neural networks (DNNs) necessitate heavy computational resource usage, such machine analytics tasks may need to be offloaded to a remote server. For example, low-end devices on the Internet of Things (IoT) record a significant amount of visual content that needs to be transmitted to a remote server to be analyzed and/or stored. To address the heavy communication requirements of such systems, new compression schemes and standards activities such as MPEG Video Coding for Machines (VCM)~\cite{vcm_cfp} have emerged, and have become attractive areas of research in recent years.

At the same time, DNN-aided data-driven image compression algorithms~\cite{balle2018variational,cheng2020image} have attracted considerable research interest as they outperform the rate-distortion (RD) performance of
off-the-shelf image codecs such as JPEG2000~\cite{christopoulos2000jpeg2000} and HEVC Intra coding~\cite{hevc_std} across various experimental setups.
Such DNN-based compression approaches are typically optimized for mean squared error (MSE) and/or multi-scale structural similarity index (MS-SSIM)~\cite{MS-SSIM}, which are used as distortion metrics between the original and reconstructed images within the loss function.
Moreover, in the case of DNN-based compression models targeting a machine vision task, the distortion metric is simply replaced with a task-specific loss, as in~\cite{torfason2018towards, chamain2021end}. 
More recently, the studies in~\cite{yan2021sssic, choi2022scalable} present DNN-based scalable compression frameworks that simultaneously support multiple tasks through scalable bitstreams sent to the decoder. For example, in~\cite{choi2022scalable}, the base layer bitstream is transmitted to the decoder for the object detection task, and the enhancement layer bitstream is additionally transmitted only when the reconstruction of input images is required. To optimize the compression performance of such a scalable system, the compression model must learn how to separate the information into different parts necessary for each task without any significant overlap.

In this work, we propose a two-task scalable image codec with a new variational formulation alternative to the current state-of-the-art proposed in \cite{choi2022scalable}.
Our scalable codec provides a base layer supporting a machine vision task, with significant gains in RD-performance compared to relevant benchmarks, and an enhancement layer supporting an input reconstruction task. 
In Section~\ref{sec:related}, we briefly summarize the most relevant prior work. Our proposed method is then described in detail in Section~\ref{sec:proposed}, followed by comprehensive experimental results with various configurations of our model detailed in Section~\ref{sec:experiments}. Furthermore, in Section~\ref{sec:info_flow}, we present an in-depth comparison of our model with the most relevant benchmark from an information-theoretic perspective. Finally, in Section~\ref{sec:conclusion} we conclude on the analyzed approaches and suggest possible future research directions.

\section{Related Work} \label{sec:related}

\begin{figure}
    \centering
    \begin{minipage}{0.22\linewidth}
        \centering
        \includegraphics[width=\textwidth]{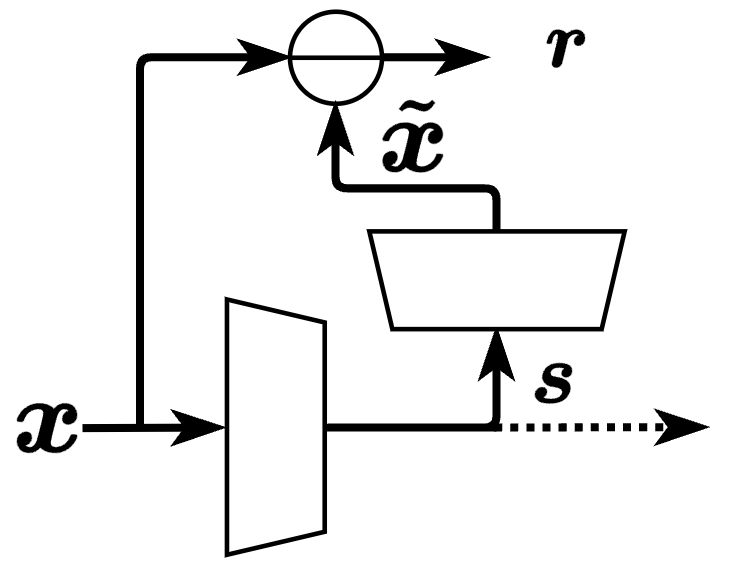}
        \centering
        \centerline{\footnotesize{(a)}}
    \end{minipage}%
    \hspace{1.2cm}%
    \begin{minipage}{0.22\linewidth}
        \centering
        \includegraphics[width=\textwidth]{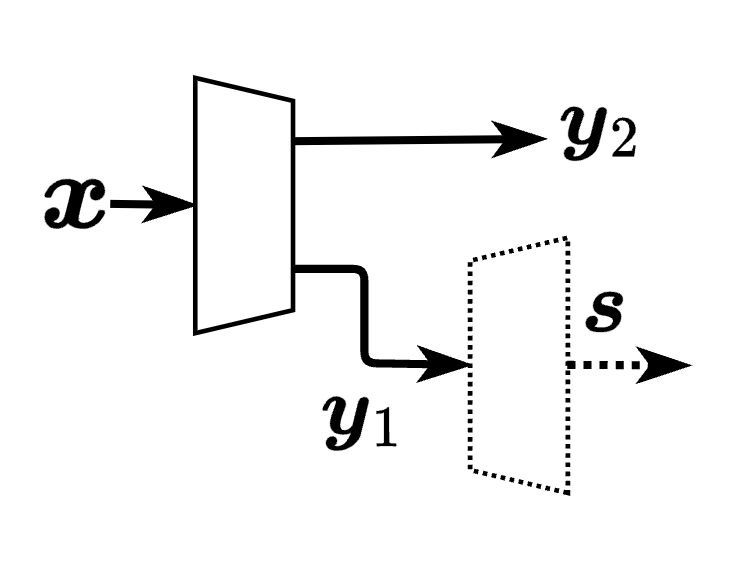}
        \centering
        \centerline{\footnotesize{(b)}}
    \end{minipage}%
    \hspace{1.2cm}%
    \begin{minipage}{0.22\linewidth}
        \centering
        \includegraphics[width=\textwidth]{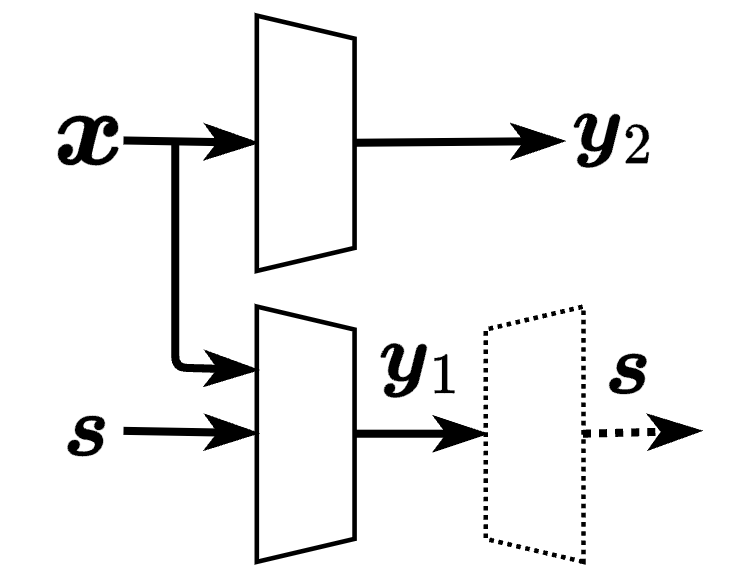}
        \centering
        \centerline{\footnotesize{(c)}}
    \end{minipage}%
\vspace{-0.2cm}
\caption{\footnotesize{Various methods to separate information into task-relevant data for multiple tasks in a scalable manner. The data separation is achieved by
(a) generating feature data $\boldsymbol{s}$ alongside a residual $\boldsymbol{r} = \boldsymbol{x} - \tilde{\boldsymbol{x}}$ which encodes the error in the decoder-side reconstruction $\tilde{\boldsymbol{x}}$ of the original input $\boldsymbol{x}$,
(b) transforming $\boldsymbol{x}$ with a single learned encoder, and
(c) transforming $\boldsymbol{x}$ and $\boldsymbol{s}$ with two learned encoders into two latent representations $\{\boldsymbol{y}_1, \boldsymbol{y}_2\}$.
Dotted lines denote the decoder operations.}}
\vspace{-0.2cm}
\label{fig:information_split}
\end{figure}

Different approaches have been explored in the literature in order to build codecs that separate the information into multiple parts associated with corresponding end tasks, some of which are shown in Fig.~\ref{fig:information_split}. For example, Yan \textit{et al.}~\cite{yan2021sssic} adopts a scheme shown in Fig.~\ref{fig:information_split}(a) in which the feature representation $\boldsymbol{s}$, designated for a vision task, is extracted from the input $\boldsymbol{x}$ with a learned transform, in order to be consecutively compressed and transmitted to the decoder. Therefore, at the decoder, $\boldsymbol{s}$ is used as an input to a computer vision network. 
Additionally, an estimate of the input $\tilde{\boldsymbol{x}}$ can be determined from $\boldsymbol{s}$ using an auxiliary module.
Using this estimate, an encoder may also compress and transmit a residual $\boldsymbol{r} = \boldsymbol{x} - \tilde{\boldsymbol{x}}$.
At the decoder side, $\boldsymbol{r}$ may be used in conjunction with the previously transmitted $\boldsymbol{s}$ in order to reconstruct the input $\boldsymbol{x}$.
However, in this scheme, the optimality of $\boldsymbol{r}$ with respect to the image reconstruction task depends on how well $\tilde{\boldsymbol{x}}$ can be reconstructed from the feature representation $\boldsymbol{s}$.

Choi \textit{et al.}~\cite{choi2022scalable} introduce a latent-space scalable codec based on the Cheng \textit{et al.}~\cite{cheng2020image} architecture.
As shown in Fig.~\ref{fig:information_split}(b), a single learned transform in the encoder maps the input $\boldsymbol{x}$ into a latent space consisting of two learned latent representations $\boldsymbol{y}_1$ and $\boldsymbol{y}_2$. To carry out the machine vision task at the decoder side, $\boldsymbol{y}_1$ is subsequently fed into another learned transform, referred to as \emph{latent-space transform}, in order to obtain an estimate of $\boldsymbol{s}$, which will be used as an input to a computer vision network. For the input reconstruction task, both latent representations $\boldsymbol{y}_1$ and $\boldsymbol{y}_2$ are concatenated and used as input to a pixel reconstruction decoder. %
In order to ensure that there is no loss in compression efficiency when coding $\boldsymbol{y}_1$ and $\boldsymbol{y}_2$ separately, the aforementioned latent representations should be maximally independent of one another.
\iffalse
Thus, a model should ideally be trained to achieve $I(\boldsymbol{y}_1;\boldsymbol{y}_2)=0$, where $I(\cdot \, ; \, \cdot)$ denotes information-theoretic mutual information (MI)~\cite{Cover_Thomas_2006}.
\else
Thus, a model should ideally be trained to minimize the information-theoretic mutual information (MI)~\cite{Cover_Thomas_2006} quantity $I(\boldsymbol{y}_1;\boldsymbol{y}_2)$.
\fi
Although works such as~\cite{MINE} propose methods for estimating MI, it is well-known that estimating MI is especially challenging for high dimensional variables.

In this study, we explore an alternative framework by introducing $\boldsymbol{s}$ as an additional input to a learned transform in the encoder, as shown in Fig.~\ref{fig:information_split}(c).
This helps reduce the size of the bitstream associated with $\boldsymbol{y}_1$, while allowing $\boldsymbol{y}_1$ and $\boldsymbol{y}_2$ to become more disentangled.
Making $\boldsymbol{y}_1$ more compact should be indeed beneficial for the VCM paradigm, which considers the vision inference as its primary task.

\section{Proposed Framework} \label{sec:proposed}
\begin{figure}
\centering
\begin{minipage}{.5\linewidth}
  \centering
  \includegraphics[width=\textwidth]{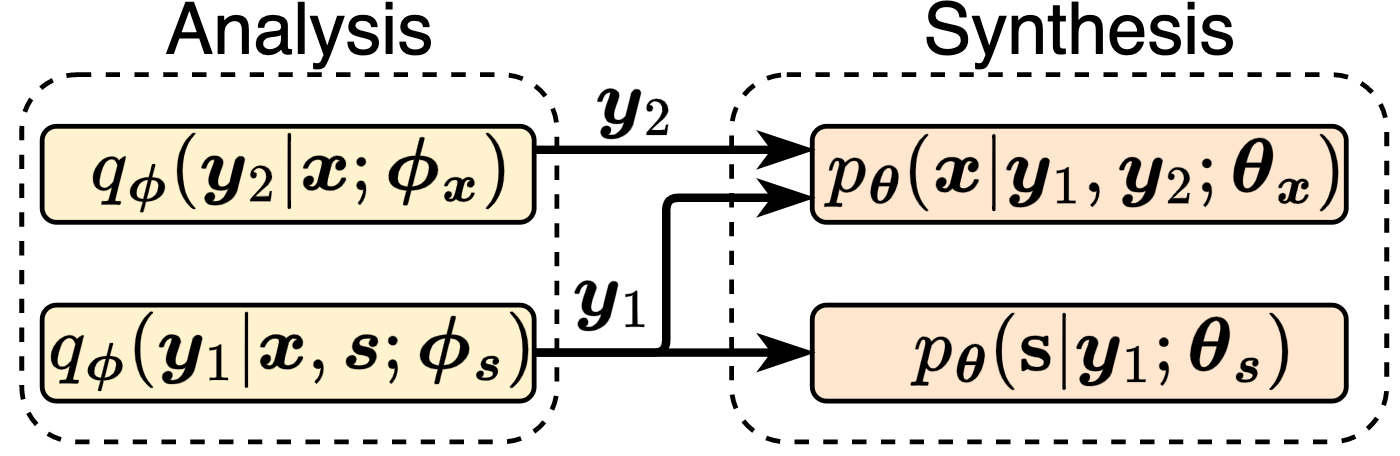}
  \centering
\end{minipage}
\vspace{-0.2cm}
\caption{\footnotesize{VAE-style compression model for our proposed method with latent-space scalability.}}
\label{fig:proposed_model}
\vspace{-0.2cm}
\end{figure}

We propose a DNN-based compression model that supports both object detection and input reconstruction tasks.
We follow a methodology similar to the two-task scalable compression model with latent-space scalability in~\cite{choi2022scalable}, but with some architectural changes.
Namely, we feed the feature representation $\boldsymbol{s}$ that is outputted by an intermediate layer of a vision task model directly as an input into our encoder. Fig.~\ref{fig:proposed_model} provides a conceptual system architecture for our proposed method, where the base latent representation $\boldsymbol{y}_1$ is designated to capture the common information between $\boldsymbol{x}$ and $\boldsymbol{s}$, whereas the enhancement latent representation $\boldsymbol{y}_2$ captures information only relevant to $\boldsymbol{x}$. To be optimal from a compression point of view, the information in $\boldsymbol{y}_2$ should not have any overlap with the information in $\boldsymbol{y}_1$. In this section, we discuss the rationale for this compression model with a variational formulation as well as the implementation details of the neural network architecture.

\subsection{Scalable compression model with an alternative variational formulation}
\label{ssec:compression_model}

Similar to~\cite{kingma}, we derive our formulation from a Bayesian variational inference perspective. Given sample observations of a random variable $x$, accompanied with a generative model $p(x \mid y)$, one seeks a posterior distribution $p(y \mid x)$. The posterior distribution cannot be, in general, expressed in closed form. Therefore, one can approximate it using a variational density of $q(y \mid x)$. One may then parameterize
the approximate posterior as
$q_{\phi}(\boldsymbol{y} \mid \boldsymbol{x}; \phi)$ and
the generative model distribution as
$p_{\theta}(\boldsymbol{x} \mid \boldsymbol{y}; \theta)$,
and consecutively, seek to minimize the Kullback--Leibler (KL) divergence
$D_{\mathrm{KL}}(
  q_{\phi}(\boldsymbol{y} \mid \boldsymbol{x}; \phi)
  \parallel
  p_{\theta}(\boldsymbol{y} \mid \boldsymbol{x}; \theta)
)$.

In our case, the distributions $q_{\phi}(\boldsymbol{y}_{1} \mid \boldsymbol{x}, \boldsymbol{s}; \phi_{\boldsymbol{s}})$ and $q_{\phi}(\boldsymbol{y}_{2} \mid \boldsymbol{x}; \phi_{\boldsymbol{x}})$ are learned by the encoder-side analysis transforms $g_{e,\boldsymbol{s}}$ and $g_{e,\boldsymbol{x}}$, respectively, and are parameterized by the weights $\phi_{\boldsymbol{s}}$ and $\phi_{\boldsymbol{x}}$.
Both latent representations $\boldsymbol{y}_{1}$ and $\boldsymbol{y}_{2}$ are then rounded to the closest integer values to obtain $\hat{\boldsymbol{y}}_{1}$ and $\hat{\boldsymbol{y}}_{2}$ before being fed into an entropy coder. During training, we follow the same strategy for quantization as in~\cite{balle2017}, by replacing the rounding operation with additive uniform random noise to obtain ``noisy'' counterparts of the latent representations, $\tilde{\boldsymbol{y}}_{1}$ and $\tilde{\boldsymbol{y}}_{2}$, which approximate $\hat{\boldsymbol{y}}_{1}$ and $\hat{\boldsymbol{y}}_{2}$.

At the decoder side, the synthesis transforms $g_{d,\boldsymbol{x}}$ and $g_{d,\boldsymbol{s}}$ learn the marginal distributions $p_{\theta}(\boldsymbol{x} \mid \boldsymbol{y}_{1}, \boldsymbol{y}_{2} ; \theta_{\boldsymbol{x}})$ and $p_{\theta}(\boldsymbol{s} \mid \boldsymbol{y}_{1}; \theta_{\boldsymbol{s}})$, both parameterized by
the weights
$\theta_{x}$ and $\theta_{s}$, respectively. Note that $\boldsymbol{y}_{1}$ is jointly learned from both variables $\boldsymbol{x}$ and $\boldsymbol{s}$, and also is given as an input to both synthesis transforms. Conversely, $\boldsymbol{y}_{2}$ is only extracted from the input image $\boldsymbol{x}$ and is thus given as input only to $g_{d,x}$.

We model the variables $\boldsymbol{y}_{1}$ and $\boldsymbol{y}_{2}$ using a parametric, fully factorized density function as in~\cite{balle2018variational}.
More specifically, each element of the latent representations
is modeled as a zero-mean Gaussian distribution with a standard deviation that is predicted from a latent via a \emph{hyperprior} block.
Following the graphical model induced in Fig.~\ref{fig:proposed_model}, we model the joint distribution of random variables as
$p_{\theta}(\boldsymbol{x}, \boldsymbol{s}, \boldsymbol{y}_{1}, \boldsymbol{y}_{2})
  = p(\boldsymbol{y}_{1})
    p(\boldsymbol{y}_{2})
    p_{\theta}(\boldsymbol{x} \mid \boldsymbol{y}_{1}, \boldsymbol{y}_{2} ; \theta_{\boldsymbol{x}})
    p_{\theta}(\boldsymbol{s} \mid \boldsymbol{y}_{1}; \theta_{\boldsymbol{s}})
$.
In order to approximate the posterior densities of the latent variables, we factorize the approximate posterior distribution as
$q_{\phi}(\boldsymbol{y}_{1}, \boldsymbol{y}_{2} \mid \boldsymbol{x}, \boldsymbol{s})
  = q_{\phi}(\boldsymbol{y}_{1} \mid \boldsymbol{x}, \boldsymbol{s}; \phi_{\boldsymbol{s}})
    q_{\phi}(\boldsymbol{y}_{2} \mid \boldsymbol{x}; \phi_{\boldsymbol{x}})
$.
Then, the loss function to minimize is given by the KL divergence
between the approximate posterior
$q_{\phi}(\tilde{\boldsymbol{y}}_{1}, \tilde{\boldsymbol{y}}_{2} \mid \boldsymbol{x}, \boldsymbol{s})$
and the true posterior
$p_{\theta}(\tilde{\boldsymbol{y}}_{1}, \tilde{\boldsymbol{y}}_{2} \mid \boldsymbol{x}, \boldsymbol{s})$
over the data distribution
$p(\boldsymbol{x}, \boldsymbol{s})$:
{\small
\begin{align}
    \mathcal{L}
    &= D_{\mathrm{KL}}(q_{\phi}(\tilde{\boldsymbol{y}}_{1}, \tilde{\boldsymbol{y}}_{2} \mid \boldsymbol{x}, \boldsymbol{s}) \parallel  p_{\theta}(\tilde{\boldsymbol{y}}_{1}, \tilde{\boldsymbol{y}}_{2} \mid \boldsymbol{x}, \boldsymbol{s}) \mid p(\boldsymbol{x}, \boldsymbol{s})) \nonumber  \\
   &= \mathbb{E}_{\boldsymbol{x}, \boldsymbol{s} \sim p(\boldsymbol{x}, \boldsymbol{s})} \Big[D_{\mathrm{KL}}\left(q_{\phi}(\tilde{\boldsymbol{y}}_{1}, \tilde{\boldsymbol{y}}_{2} \mid \boldsymbol{x}, \boldsymbol{s}) \parallel  p_{\theta}(\tilde{\boldsymbol{y}}_{1}, \tilde{\boldsymbol{y}}_{2} \mid \boldsymbol{x}, \boldsymbol{s})\right) \Big] \nonumber \\
   &= \mathbb{E}_{\boldsymbol{x}, \boldsymbol{s} \sim p(\boldsymbol{x}, \boldsymbol{s})} \mathbb{E}_{\tilde{\boldsymbol{y}}_{1}, \tilde{\boldsymbol{y}}_{2} \sim q_{\phi}}\Big[ \Big(\log q_{\phi}(\tilde{\boldsymbol{y}}_{1} \mid \boldsymbol{x}, \boldsymbol{s}; \phi_{\boldsymbol{s}}) + \log q_{\phi}(\tilde{\boldsymbol{y}}_{2} \mid \boldsymbol{x}; \phi_{\boldsymbol{x}}) \Big)  \nonumber \\
   & - \Big( \underbrace{\log p_{\theta}(\boldsymbol{x} \mid \tilde{\boldsymbol{y}}_{1}, \tilde{\boldsymbol{y}}_{2} ; \theta_{\boldsymbol{x}})}_{D_{\boldsymbol{x}}} +  \underbrace{\log p_{\theta}(\boldsymbol{s} \mid \tilde{\boldsymbol{y}}_{1};\theta_{\boldsymbol{s}})}_{D_{\boldsymbol{s}}} + \underbrace{\log p(\tilde{\boldsymbol{y}}_{1})}_{R_{y_{1}}} + \underbrace{\log p(\tilde{\boldsymbol{y}}_{2})}_{R_{\boldsymbol{y}_{2}}} \Big)\Big] + \text{const.}
   \label{eq:KL}
\end{align}
}%
The first parenthesized group of terms within the expectation is zero since the densities $q$ are a product of uniform densities of unit width, due to perturbation with uniform noise during training.
The terms labeled $D_x$ and $D_s$ coincide with distortion terms associated with the input image $\boldsymbol{x}$ and feature representation $\boldsymbol{s}$ for the targeted vision task, respectively.
The terms labeled $R_{y_1}$ and $R_{y_2}$ represent the cross-entropy values, corresponding to the bit costs of encoding $\tilde{\boldsymbol{y}}_{1}$ and $\tilde{\boldsymbol{y}}_{2}$ under the respective learned entropy models.

By linking the parameterized density functions to the transform coding paradigm, we observe that the minimization of the KL divergence effectively corresponds to optimizing the VCM model for rate-distortion performance associated with both input image reconstruction and object detection tasks. In the case of using the MSE metric, the distortion terms $D_{\boldsymbol{x}}$ and $D_{\boldsymbol{s}}$ in Eq.~\eqref{eq:KL} correspond to closed-form likelihoods, or more specifically, to Gaussian distributions (see \cite{balle2017} for relevant discussion). We can write the loss function from Eq.~\eqref{eq:KL} compactly as
{\small
\begin{equation}
    \mathcal{L}  = R_{\boldsymbol{y}_{1}} + R_{\boldsymbol{y}_{2}} + \lambda \cdot D_{\boldsymbol{x}} + \gamma \cdot D_{\boldsymbol{s}},
    \label{eq:compact}
\end{equation}}%
where the scalars $\lambda$ and $\gamma$ denote the Lagrange multipliers corresponding to the distortion budgets associated with $\boldsymbol{x}$ and $\boldsymbol{s}$, respectively.

\subsection{Implementation of the neural network architecture}
\label{ssec:implemtned_architecture}

\begin{figure}
\centering
\begin{minipage}{0.85\linewidth}
  \centering
  \includegraphics[width=\textwidth]{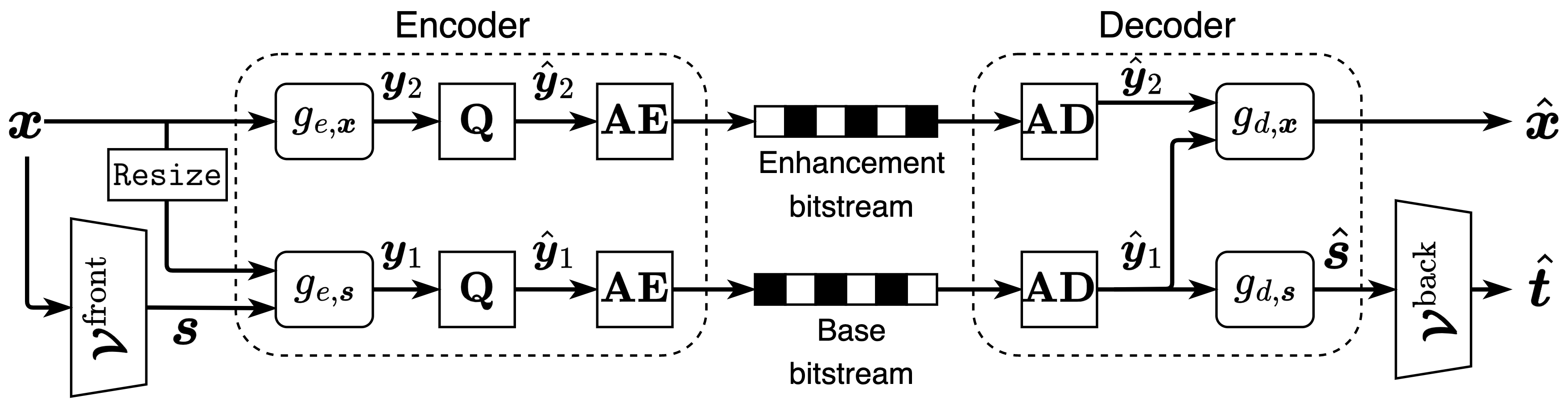}
  \centering
\end{minipage}%
\caption{\footnotesize{Schematic of the proposed neural network architecture. 
Hyperprior blocks and side information bitstreams similar to~\cite{balle2018variational} are also present, but are not visualized here.}}
\label{fig:schematic_architecture}
\vspace{-0.2cm}
\end{figure}

\begin{table}[t]
\centering
\caption{Network layer configurations of the encoder and of the decoder.}
\vspace{-0.2cm}
\label{tbl:details_enc_dec}
\smallskip\noindent
\resizebox{1.0\linewidth}{!}{%
\renewcommand{\arraystretch}{0.5}
\begin{tabular}{@{}c|cccc|cccc@{}}
\toprule
\multirow{2}{*}{} & \multicolumn{4}{c|}{Encoder}                                            & \multicolumn{4}{c}{Decoder}                                              \\ \cmidrule(l){2-9} 
                  & \multicolumn{2}{c|}{$g_{e,\boldsymbol{s}}$}               & \multicolumn{2}{c|}{$g_{e,\boldsymbol{x}}$} & \multicolumn{2}{c|}{$g_{d,\boldsymbol{s}}$}                 & \multicolumn{2}{c}{$g_{d,\boldsymbol{x}}$} \\ \midrule
No.               & Layer   & \multicolumn{1}{c|}{In/Out} & Layer       & In/Out    & Layer     & \multicolumn{1}{c|}{In/Out} & Layer       & In/Out   \\ \midrule
1                 & conv5s1 & \multicolumn{1}{c|}{$C_{\boldsymbol{s}}+3/N$}    & conv5s2     & $3/N$           & deconv5s1 & \multicolumn{1}{c|}{$M_1/N$}    & deconv5s2   & $M/N$         \\
2                 & conv5s1 & \multicolumn{1}{c|}{$N/N$}       & conv5s2     & $N/N$          & deconv5s1 & \multicolumn{1}{c|}{$N/N$}       & deconv5s2   & $N/N$         \\
3                 & conv5s2 & \multicolumn{1}{c|}{$N/M_1$}    & conv5s2     & $N/N$          & deconv5s2 & \multicolumn{1}{c|}{$N/C_{\boldsymbol{s}}$}      & deconv5s2   & $N/N$         \\
4                 &         & \multicolumn{1}{c|}{}           & conv5s2     & $N/M_2$       &           & \multicolumn{1}{c|}{}           & deconv5s2   & $N/3$         \\ \bottomrule
\end{tabular}}
\vspace{-0.2cm}
\end{table}

As seen in Fig.~\ref{fig:schematic_architecture}, we build our neural network architecture based on the approach in~\cite{balle2018variational}.
We generate the feature representation $\boldsymbol{s} \in \mathbb{R}^{C_{\boldsymbol{s}} \times H_{\boldsymbol{s}} \times W_{\boldsymbol{s}}}$ by feeding the input image $\boldsymbol{x} \in \mathbb{R}^{3\times H \times W}$ through the first few layers of a machine vision model, denoted as $\mathbfcal{V}^{\textup{front}}$ in Fig.~\ref{fig:schematic_architecture}.
To have a fair comparison with~\cite{choi2022scalable}, we use the first consecutive 13 layers of the YOLOv3~\cite{yolov3} object detection model to generate $\boldsymbol{s}$ with
$C_s = 256$ channels, and spatial dimensions of $H_s = \frac{H}{8}$ and $W_s = \frac{W}{8}$.

The analysis transform $g_{e,\boldsymbol{s}}$ generates the base latent representation $\boldsymbol{y}_1$ using as input the channel-wise concatenation of $\boldsymbol{s}$ and $\Resize(\boldsymbol{x})$,
where we have chosen a spatial bicubic interpolation filter for $\Resize : \mathbb{R}^{3\times H \times W} \to \mathbb{R}^{3 \times H_{\boldsymbol{s}} \times W_{\boldsymbol{s}}}$.
The analysis transform $g_{e,\boldsymbol{x}}$ generates the enhancement latent representation $\boldsymbol{y}_2$ using only $\boldsymbol{x}$ as input.
Next, $\boldsymbol{y}_1$ and $\boldsymbol{y}_2$
are quantized ($\mathbf{Q}$) and fed into an arithmetic encoder $(\mathbf{AE})$, which yields the base and enhancement bitstreams, respectively.

At the decoder side, the respective bitstreams are then fed into an arithmetic decoder $(\mathbf{AD})$ in order to reconstruct the base and enhancement latent representations $\hat{\boldsymbol{y}}_1$ and $\hat{\boldsymbol{y}}_2$. Using $\hat{\boldsymbol{y}}_1$, the synthesis transform $g_{d,\boldsymbol{s}}$ reconstructs the feature representation $\hat{\boldsymbol{s}}$, which we feed into the remaining part of the machine vision model, denoted as $\mathbfcal{V}^{\textup{back}}$, in order to generate the inference results $\hat{\boldsymbol{t}}$. Using the channel-wise concatenation of $\hat{\boldsymbol{y}}_1$ and $\hat{\boldsymbol{y}}_2$, the synthesis transform $g_{d,\boldsymbol{x}}$ reconstructs the input $\hat{\boldsymbol{x}}$.

As our network architecture builds upon~\cite{balle2018variational}, it employs separate hyperprior modules for both latent representations $\boldsymbol{y}_1$ and $\boldsymbol{y}_2$. However, these are omitted in Fig.~\ref{fig:schematic_architecture} for brevity.
The layer configurations for the hyperprior modules are the same as those presented in~\cite{balle2018variational}, whereas details on the employed encoder/decoder modules are shown in Table~\ref{tbl:details_enc_dec}.
We adopt a similar configuration for the encoder/decoder architecture as in~\cite{balle2018variational}, for our $g_{e,i}$ and $g_{d,i}$ where $i=\{\boldsymbol{x}, \boldsymbol{s}\}$.
We define the analysis transforms $g_{e,i}$ using convolutional layers with $5\times5$ kernels and a stride of~2 (i.e., conv5s2), interleaved with generalized divisive normalization (GDN) layers~\cite{balle2015density}.
The synthesis transforms $g_{d,i}$ consist of transposed convolutional layers
for upsampling with a stride of 2 (i.e.,~deconv5s2), interleaved with inverse GDN (IGDN) layers.
Note that in order to match the spatial dimension of the latent representations, the number of layers at the analysis/synthesis transforms both at the encoder and decoder sides differ.
Table~\ref{tbl:details_enc_dec} lists the layers used in our model, along with their corresponding input and output channel dimensions.
In our experiments, we fix $N=192$ for all models, and vary $M_1$ and $M_2$ depending on the configuration as detailed in Section~\ref{ssec:results}.

\section{Experiments and Results} \label{sec:experiments}

\subsection{Experimental setup}

We implemented all DNN-based models using the CompressAI library~\cite{compressai}.
The models are trained on randomly cropped patches of size $256 \times 256$ from the Vimeo-90K \cite{vime90k} dataset, with a batch size of $8$.
We use an Adam optimizer with an initial learning rate of $1 \times 10^{-4}$, where the rate is reduced by a factor of $10$ whenever the decrease in validation loss stagnates, up to $4$ times, after which we stop training.
We use the loss function from Eq.~\ref{eq:compact} with $D_{\boldsymbol{x}}=\textup{MSE}(\boldsymbol{x}, \hat{\boldsymbol{x}})$ and $D_{\boldsymbol{s}}=\textup{MSE}(\boldsymbol{s}, \hat{\boldsymbol{s}})$.
Our models have been trained to operate across a wide range of bit rates by varying the hyperparameter
$\lambda \in \{0.0067, 0.0100, 0.0130, 0.0250, 0.0300, 0.0483\}$
and fixing $\gamma=0.006 \cdot \lambda$.

To explore how the overall performance changes with our proposed approach under various configurations, we vary the number of channels of the base and enhancement latent representations (i.e., $M_1$ and $M_2$, respectively), and the number of hyperprior blocks employed (i.e., $H$). We use the tuple $(M_1, M_2, H)$ to express each configuration.

Because our proposed approach is built upon~\cite{balle2018variational} for the sake of reduced computational load,
we have reimplemented~\cite{choi2022scalable} on top of a comparable base architecture with its original configuration $(128,64,1)$ in order to ensure a fairer comparison. We also compare with the latest standard codecs in intra-only mode using the reference implementations of HEVC\footnote{\url{http://hevc.hhi.fraunhofer.de/svn/svn_HEVCSoftware/tags/HM-16.20+SCM-8.8/}} and VVC\footnote{\url{https://vcgit.hhi.fraunhofer.de/jvet/VVCSoftware_VTM/-/tags/VTM-12.3}.} over the quantization parameters $\textup{QP}\in\{22, 25, 28, \ldots, 40\}$.

We evaluate object detection performance on COCO 2014 validation data\-set~\cite{coco2014} consisting of around 5000 JPEG-compressed images with annotated bounding boxes belonging to 80 object categories. We resize the input images to $512\times512$ with a bilinear interpolation filter before encoding.
Additionally, we also evaluate input reconstruction performance on all 24 images from the Kodak dataset~\cite{kodak_dataset}.

\begin{figure}
    \centering
    \begin{minipage}{0.31\linewidth}
        \centering
        \includegraphics[width=\textwidth]{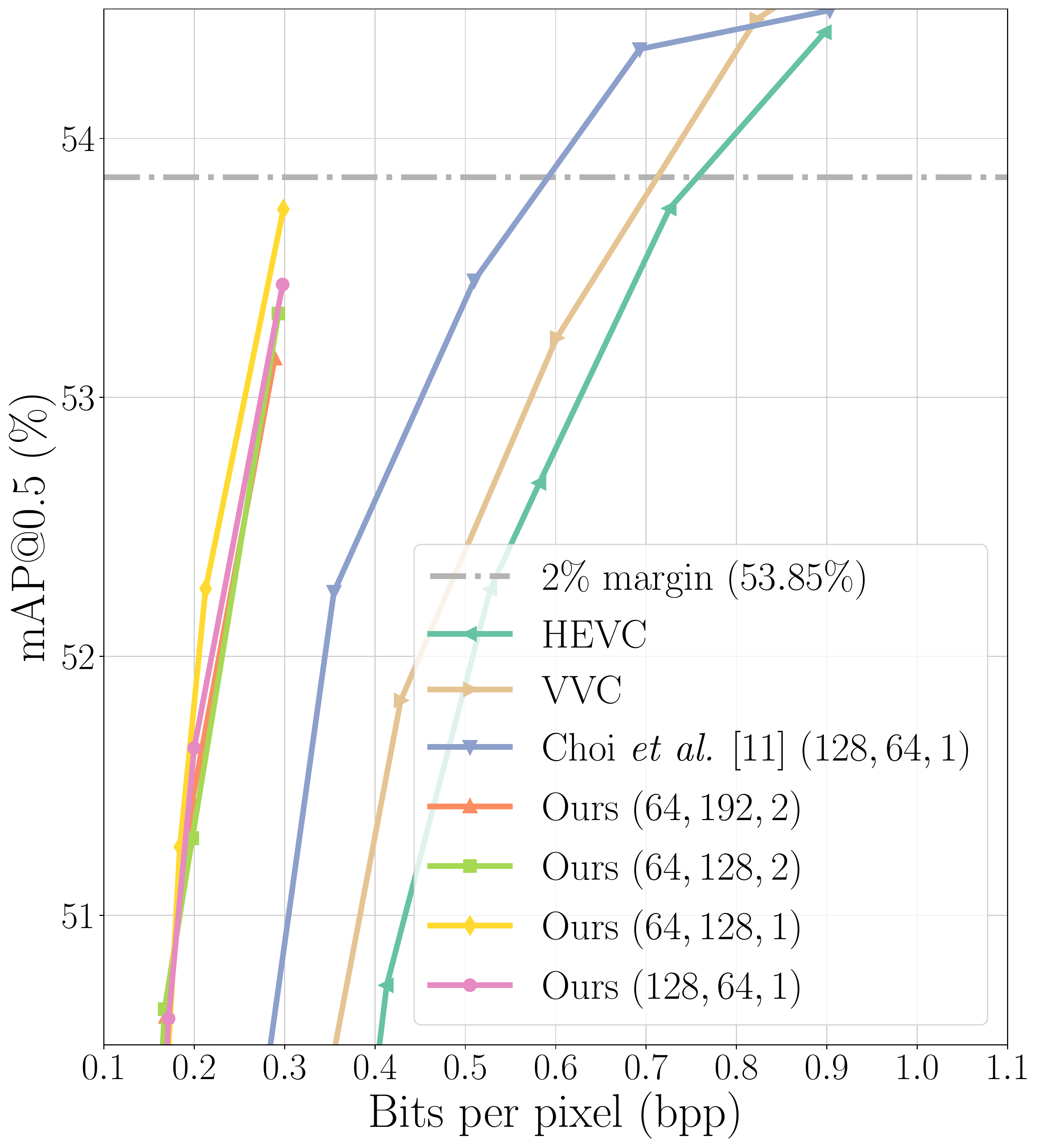}
        \centering
        \centerline{\footnotesize{(a)}}
    \end{minipage}%
    \hspace{0.02cm}
    \begin{minipage}{0.31\linewidth}
        \centering
        \includegraphics[width=\textwidth]{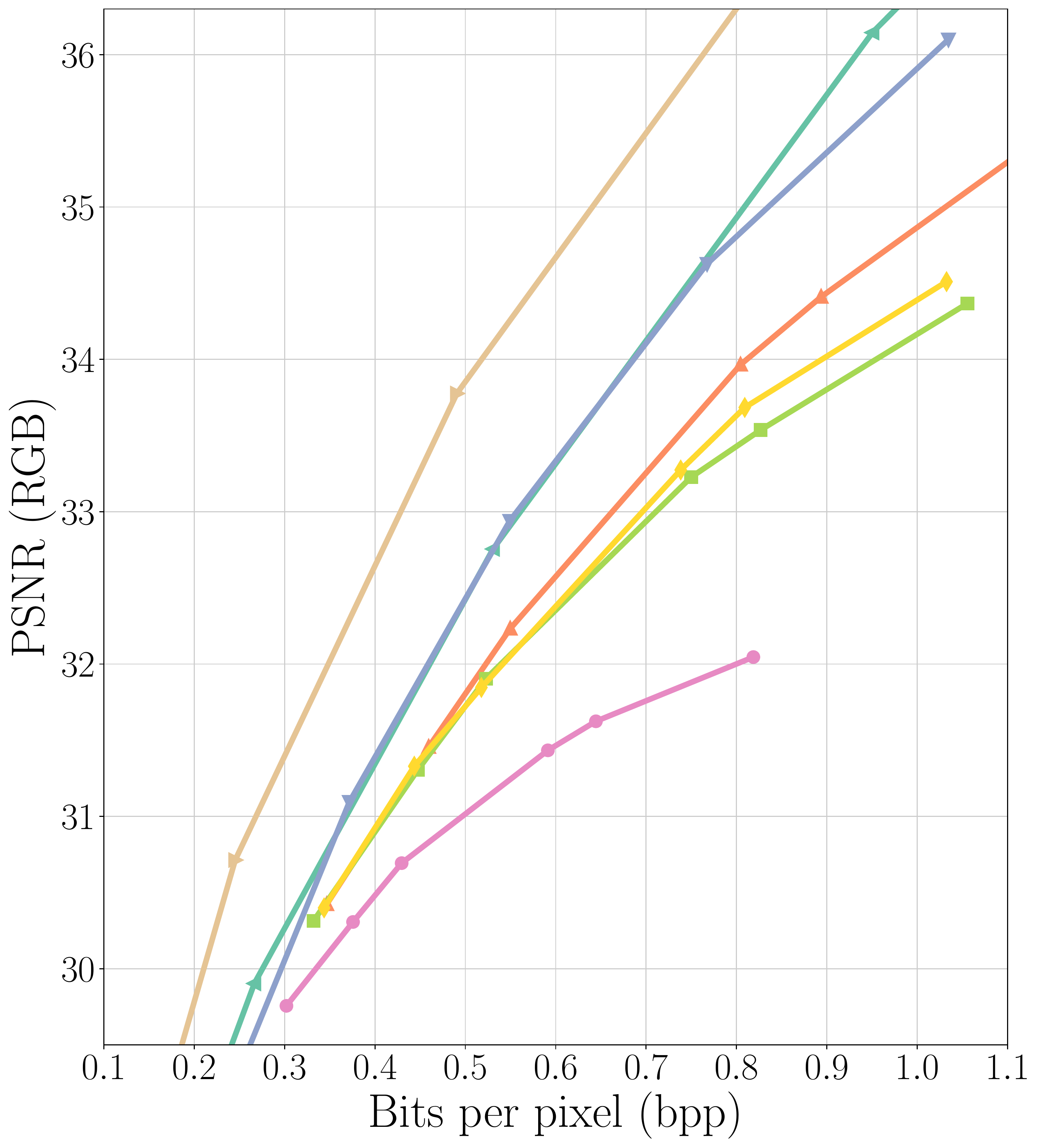}
        \centering
        \centerline{\footnotesize{(b)}}
    \end{minipage}%
    \hspace{0.02cm}
    \begin{minipage}{0.322\linewidth}
        \centering
        \includegraphics[width=\textwidth]{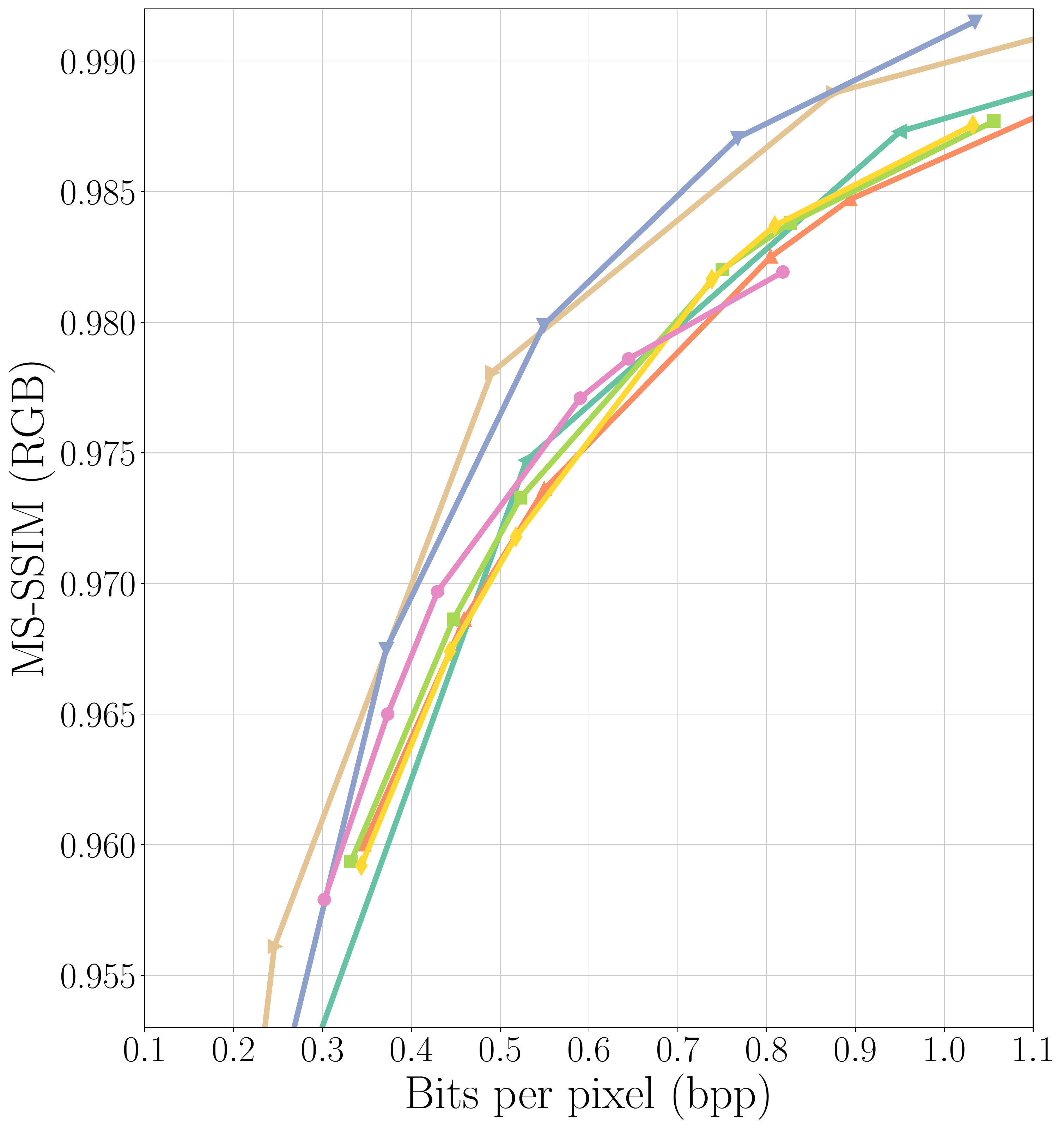}
        \centering
        \centerline{\footnotesize{(c)}}
    \end{minipage}%
\vspace{-0.2cm}
\caption{\footnotesize{Performance comparisons across various metrics. (a) Object detection in terms of mAP (IoU=0.5) vs. bpp on the COCO 2014 validation dataset. (b) Input reconstruction in terms of PSNR vs. bpp and (c) input reconstruction in terms of MS-SSIM vs. bpp on the Kodak dataset.}}
\vspace{-0.2cm}
\label{fig:performance_curves}
\end{figure}

\subsection{Results}
\label{ssec:results}

Fig.~\ref{fig:performance_curves} compares the performance of our models and relevant codecs.
Fig.~\ref{fig:performance_curves}(a) shows the object detection performance using a rate-accuracy curve in terms of mean average precision (mAP) for an Intersection over Union (IoU) threshold of 0.5 versus bits per pixel (bpp).
Fig.~\ref{fig:performance_curves}(b) and (c) show the input reconstruction performance using rate-distortion curves in terms of peak signal-to-noise ratio (PSNR) and MS-SSIM versus bpp, respectively.

For the primary machine vision task,
the object detection performance of our method with $(64, 128, 1)$ reaches near 2\% mAP loss%
\footnote{Default performance of YOLOv3 on COCO2014 dataset, including JPEG-compressed images, is around 55.85\% mAP at 4.80 bpp.}
(dashed line) at around 0.3 bpp, whereas
Choi \textit{et al.}~\cite{choi2022scalable} with a configuration of $(128, 64, 1)$
reaches a similar accuracy at around 0.58 bpp.
When repurposed for this compression task, HEVC and VVC show relatively poor performance.
In comparison with~\cite{choi2022scalable}, our method approximately achieves 55\% bit savings at the 2\% mAP loss threshold.

For the secondary input reconstruction task,
VVC achieves the best performance among all methods for both the PSNR and MS-SSIM metrics. However, our method still  shows competitive performance at low bpp compared to HEVC and Choi \textit{et al.}~\cite{choi2022scalable}.
Nonetheless, the performance gap between our method for well-chosen configurations and the benchmarks 
increases somewhat for larger bpp.
The best configuration for our method in terms of PSNR is with $M_2=192$ channels for the enhancement layer.
When comparing input reconstruction performance using MS-SSIM, all tested configurations of our method show competitive performance with respect to HEVC. 
We note that our worst-performing configuration ($M_2=64$) in terms of PSNR is still competitive when measured with MS-SSIM.

In summary, our proposed approach is capable of achieving a significant reduction in bit rate for the object detection task at the cost of slight performance degradation for the input reconstruction task.

\section{Insight into Information Flow}
\label{sec:info_flow}

\begin{figure}
    \centering
    \begin{minipage}{0.47\linewidth}
        \centering
        \includegraphics[width=\textwidth]{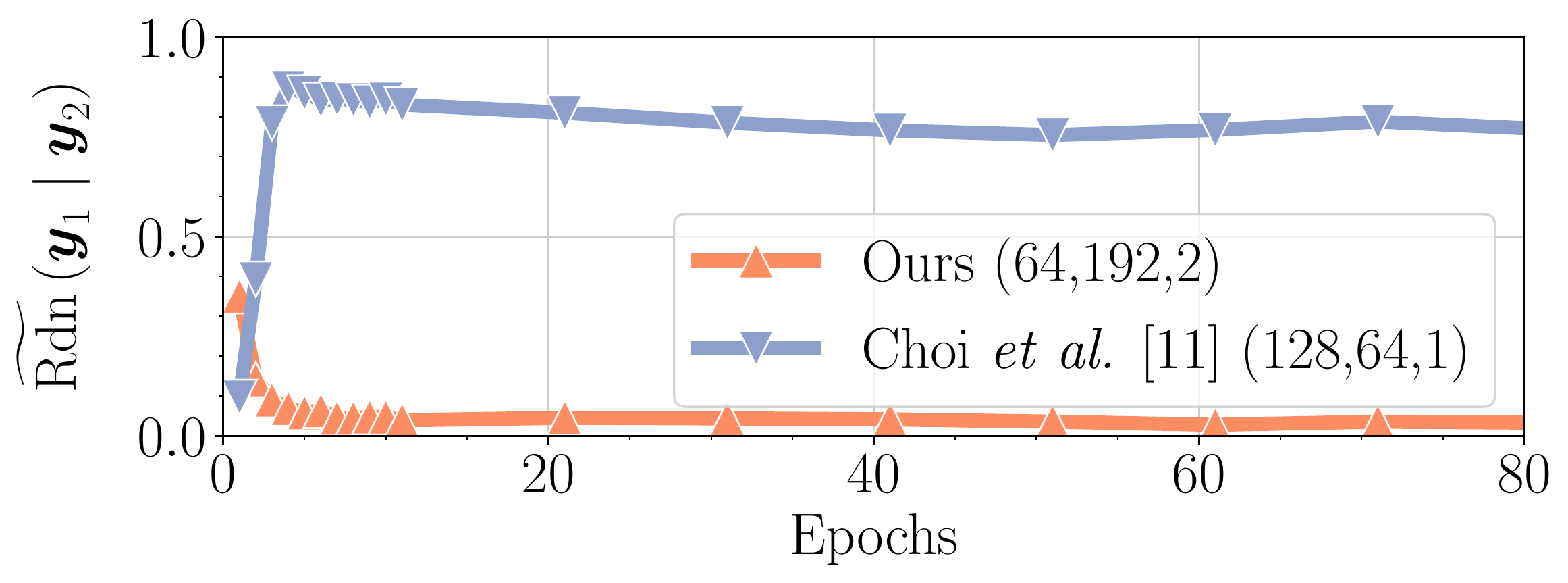}
        \centering
        \centerline{\footnotesize{(a)}}
    \end{minipage}%
    \hspace{0.04\linewidth}%
    \begin{minipage}{0.47\linewidth}
        \centering
        \includegraphics[width=\textwidth]{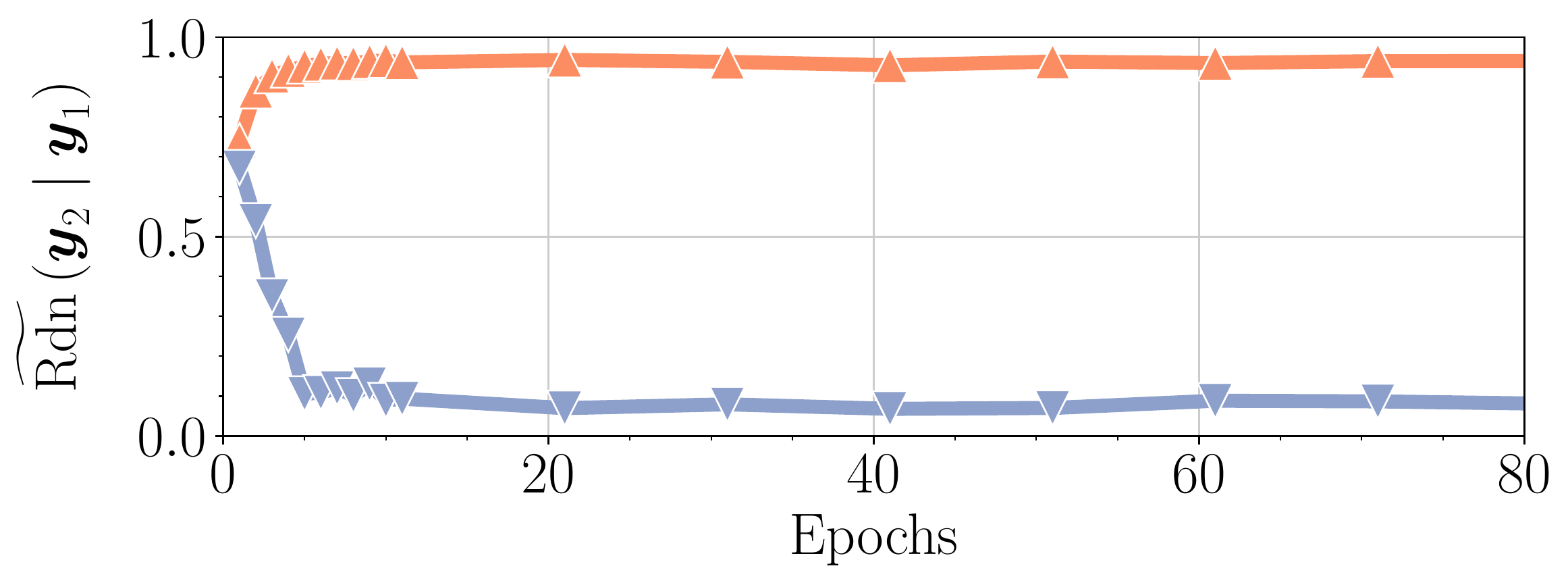}
        \centering
        \centerline{\footnotesize{(b)}}
    \end{minipage}%
\vspace{-0.2cm}
\caption{\footnotesize{Evolution during training of the redundancy metrics (a) $\EstimatedRedundancyMetric{1}{2}$ and (b) $\EstimatedRedundancyMetric{2}{1}$ discussed in Sec.~\ref{sec:info_flow}. Both curves in all figures correspond to the models trained with $\lambda=0.0483$. 
}}
\vspace{-0.2cm}
\label{fig:entp_ratio}
\end{figure}

\begin{figure}
    \centering
    \begin{minipage}{0.29\linewidth}
        \centering
        \includegraphics[width=\textwidth]{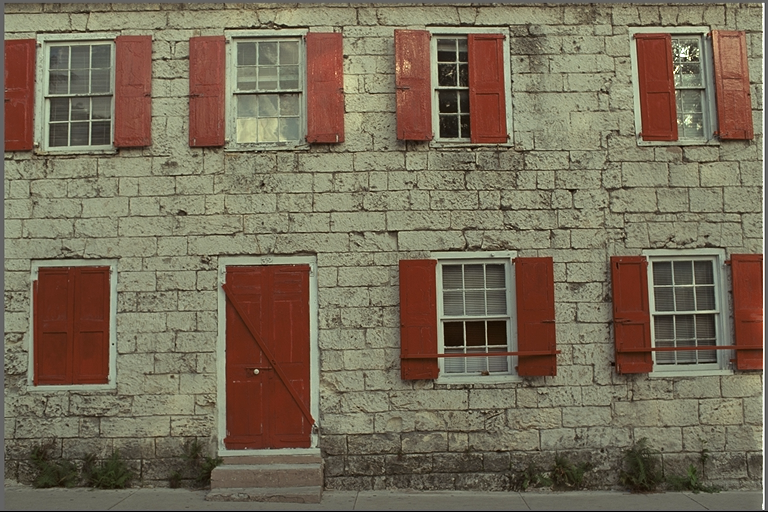}
        \centering
        \centerline{\footnotesize{(a)}}
    \end{minipage}%
    \hspace{0.02\linewidth}%
    \begin{minipage}{0.69\linewidth}
        \centering
        \includegraphics[width=\textwidth]{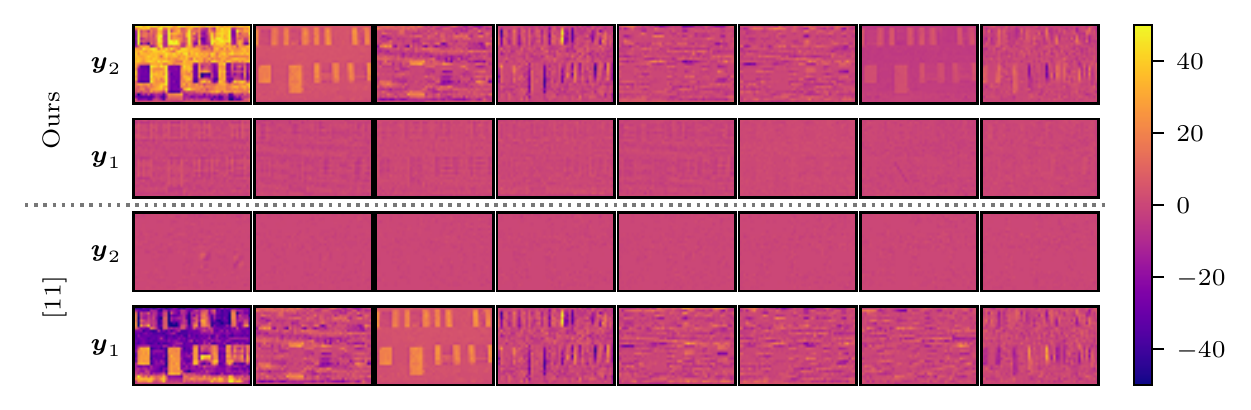}
        \centering
        \centerline{\footnotesize{(b)}}
    \end{minipage}%
\vspace{-0.2cm}
\caption{\footnotesize{(a) A sample input image from Kodak~\cite{kodak_dataset} and (b) top-8 latent channels ordered by rate for the base ($\boldsymbol{y}_{1}$) and enhancement ($\boldsymbol{y}_{2}$) latent representations of the models employed in Fig.~\ref{fig:entp_ratio}.}}
\vspace{-0.2cm}
\label{fig:samples}
\end{figure}

\iffalse
To characterize the amount of redundancy present between the base latent representation 
$\boldsymbol{y}_1$ and the enhancement latent representation 
$\boldsymbol{y}_2$,
we introduce metrics consisting of the entropy ratios
$
\FullEntropyRatio{1}{2}$ and
$
\FullEntropyRatio{2}{1}$.
If 
$\boldsymbol{y}_{1}$ 
were to contain no information about 
$\boldsymbol{y}_{2}$, 
then $\EntropyRatio{1}{2}$ would be equal to one, and vice versa for 
$\EntropyRatio{2}{1}$.
Following the conditional entropy estimation approach employed in \cite{choi2022scalable}, we separate $\boldsymbol{y}_1$ and $\boldsymbol{y}_2$ into fibers with a size of $M_m \times 1 \times 1$, where $M_m \in \{M_1, M_2\}$ is the number of channels of the respective latent tensor.
Then, we group the fibers into $K$ clusters using the $k$-means algorithm.
Finally, we compute the conditional entropy as
{\small
\begin{equation}
    H(\boldsymbol{y}_{i} \mid \boldsymbol{y}_{j}) \; \;
    \triangleq \sum_{k \in \{1, \ldots, K\}} p(k)\,H(\boldsymbol{y}_{i} \mid c(\boldsymbol{y}_{j}) = k),
\end{equation}
}%

\else

We introduce the redundancy metric $\RedundancyMetric{i}{j} \triangleq \MIRatio{i}{j} = 1 - \FullEntropyRatio{i}{j}$, also referred to as the uncertainty coefficient~\cite{theil1972statistical} in the literature, which measures what portion of the information within $\boldsymbol{y}_i$ is redundantly contained in the other variable $\boldsymbol{y}_j$. Following the conditional entropy estimation approach employed in \cite{choi2022scalable}, we separate $\boldsymbol{y}_i$ and $\boldsymbol{y}_j$ into fibers with a size of $m \times 1 \times 1$, where $m \in \{M_i, M_j\}$ is the number of channels of the respective latent tensor.
Then, we group the fibers for $\boldsymbol{y}_j$ into $K = 128$ clusters using the $k$-means algorithm.
Finally, we estimate
{\small
\begin{equation}
  \EstimatedRedundancyMetric{i}{j}
    =
    1 - \frac{1}{H(\boldsymbol{y}_i)}
    \sum_{k \in \{1, \ldots, K\}} p(k) 
    \, H(\boldsymbol{\bar{y}}_{i} \mid c(\boldsymbol{\bar{y}}_{j}) = k),
\end{equation}
}%
\fi%
where $c : \mathbb{R}^{M_j \times 1 \times 1} \rightarrow \{1, \ldots, K\}$ is a fixed clustering function,
$p(k)$ denotes the approximate probability density associated with each cluster $k$,
and $(\boldsymbol{\bar{y}}_i, \boldsymbol{\bar{y}}_j)$ is a random variable representing one of $L$ pairs of fibers sampled over 256 images.
To estimate 
$H(\boldsymbol{y}_i)$, we employ the entropy bottleneck module of~\cite{balle2018variational},
and also use it in computing the
estimate
$H(\boldsymbol{\bar{y}}_{i} \mid c(\boldsymbol{\bar{y}}_{j}) = k) \approx \sum_{l \in \{1, \ldots, L\}} H(\boldsymbol{\bar{y}}_{i}^{(l)}) \; \delta[c(\boldsymbol{\bar{y}}_{j}^{(l)}) - k]$.

We compare the evolution of the aforementioned metric during training for our method and for the benchmark model~\cite{choi2022scalable}.
As Fig.~\ref{fig:entp_ratio}(a) shows,
$\EstimatedRedundancyMetric{1}{2}$ stabilizes near the desired value of zero for our method, whereas it is larger for~\cite{choi2022scalable}.
Conversely, as Fig.~\ref{fig:entp_ratio}(b) shows, $\EstimatedRedundancyMetric{2}{1}$ stabilizes near one for our method, and zero for~\cite{choi2022scalable}.
This confirms that our proposed approach yields a more compact base latent representation, while producing a suboptimal enhancement latent representation.
Furthermore, it affirms that the model from~\cite{choi2022scalable} offers a more graceful degradation in the context of image reconstruction quality as its enhancement latent representation contains less redundancy compared to ours.
Although the loss in coding efficiency due to scalability has been previously studied in~\cite{choi2022scalable}, we argue that our way of looking into information flow through an information-theoretic lens provides deeper understanding about degree of redundancy between the latent representations.

Fig.~\ref{fig:samples} visualizes the top-8 channels, ordered by rate, of the base and enhancement latent representations for both our method and the one in~\cite{choi2022scalable}.
For our approach, $\boldsymbol{y}_1$ contains very little visible image structure, suggesting that it is well optimized for the object detection task.
Without achieving comparative gains in task accuracy, as seen in Fig.~\ref{fig:performance_curves}, \cite{choi2022scalable} produces significant visible image structure within $\boldsymbol{y}_1$, leading to a significantly higher bit cost for the base bitstream.
Thus, our method efficiently encodes only what is necessary for a given task within its respective bitstream.

\section{Conclusion}
\label{sec:conclusion}

This paper presents a DNN-based image codec with a new variational formulation, offering latent-space scalability for human and machine vision tasks by disentangling the learned latent representations. For this end, the information related to the object detection task is extracted at the encoder side to be used as an additional input, together with original input image, to a learned transform at the encoder. As such, compared to the state-of-the-art benchmark in~\cite{choi2022scalable}, we achieve significant bit reductions at the base layer bitstream for the object detection task, hence yielding a desirable scalable image codec for the VCM paradigm. Additionally, we introduce an information-theoretic metric to analyze the characteristics of the amount of redundancy between two learned latent representations. We leave the investigation of how to further improve image reconstruction quality while not compromising the object detection performance for future work.

\Section{References}
\bibliographystyle{IEEEbib-abbrev}
\bibliography{refs}

\begin{thebibliography}{10}

\bibitem{CISCO_VNI_2023}
``Cisco annual internet report (2018-2023) {W}hite paper,''
  \url{https://www.cisco.com/c/en/us/solutions/collateral/executive-perspectives/annual-internet-report/white-paper-c11-741490.html}.

\bibitem{vcm_cfp}
{Int. Standards Org./Int/Electrotech. Commun.},
\newblock ``{Call for Proposals on Video Coding for Machines},'' ISO/IEC JTC
  1/SC 29/WG 2/N220, July 2022.

\bibitem{balle2018variational}
J. Ball{\'e}, D. Minnen, S. Singh, S.~J. Hwang, and N. Johnston,
\newblock ``Variational image compression with a scale hyperprior,''
\newblock in {\em Proc. ICLR}, 2018.

\bibitem{cheng2020image}
Z. Cheng, H. Sun, M. Takeuchi, and J. Katto,
\newblock ``Learned image compression with discretized gaussian mixture
  likelihoods and attention modules,''
\newblock in {\em Proc. IEEE CVPR}, 2020.

\bibitem{christopoulos2000jpeg2000}
C. Christopoulos, A. Skodras, and T. Ebrahimi,
\newblock ``The {JPEG}2000 still image coding system: {A}n overview,''
\newblock {\em IEEE Trans. Consum. Electron.}, vol. 46, no. 4, pp. 1103--1127,
  2000.

\bibitem{hevc_std}
{Int. Telecommun. Union-Telecommun. and Int. Standards Org./Int/Electrotech.
  Commun.},
\newblock ``High efficiency video coding,'' Rec. ITU-T H.265 and ISO/IEC
  23008-2, 2019.

\bibitem{MS-SSIM}
Z. Wang, E. Simoncelli, and A. Bovik,
\newblock ``Multiscale structural similarity for image quality assessment,''
\newblock in {\em Proc. Asilomar Conf. Signals, Systems, and Computers}, 2003.

\bibitem{torfason2018towards}
R. Torfason, F. Mentzer, E. Agustsson, M. Tschannen, R. Timofte, and L.
  Van~Gool,
\newblock ``Towards image understanding from deep compression without
  decoding,''
\newblock {\em arXiv preprint arXiv:1803.06131}, 2018.

\bibitem{chamain2021end}
L.~D. Chamain, F. Racap{\'e}, J. B{\'e}gaint, A. Pushparaja, and S. Feltman,
\newblock ``End-to-end optimized image compression for machines, a study,''
\newblock in {\em Proc. IEEE DCC}, 2021, pp. 163--172.

\bibitem{yan2021sssic}
N. Yan, C. Gao, D. Liu, H. Li, L. Li, and F. Wu,
\newblock ``{SSSIC}: semantics-to-signal scalable image coding with learned
  structural representations,''
\newblock {\em IEEE Trans. Image Processing}, vol. 30, pp. 8939--8954, 2021.

\bibitem{choi2022scalable}
H. Choi and I.~V. Baji{\'c},
\newblock ``Scalable image coding for humans and machines,''
\newblock {\em IEEE Trans. Image Processing}, vol. 31, pp. 2739--2754, 2022.

\bibitem{Cover_Thomas_2006}
T.~M. Cover and J.~A. Thomas,
\newblock {\em Elements of Information Theory},
\newblock Wiley, 2nd edition, 2006.

\bibitem{MINE}
M.~I. Belghazi, A. Baratin, S. Rajeswar, S. Ozair, Y. Bengio, A. Courville, and
  R.~D. Hjelm,
\newblock ``{MINE}: {M}utual information neural estimation,''
\newblock {\em arXiv preprint arXiv:1801.04062}, 2018.

\bibitem{kingma}
D.~P. Kingma and M. Welling,
\newblock ``Auto-encoding variational bayes,''
\newblock {\em arXiv preprint arXiv:1312.6114}, 2013.

\bibitem{balle2017}
J. Ball{\'e}, V. Laparra, and E.~P. Simoncelli,
\newblock ``End-to-end optimized image compression,''
\newblock in {\em Proc. ICLR}, 2017.

\bibitem{yolov3}
J. Redmon and A. Farhadi,
\newblock ``{YOLOv3}: {A}n incremental improvement,''
\newblock {\em arXiv preprint arXiv:1804.02767}, 2018.

\bibitem{balle2015density}
J. Ball{\'e}, V. Laparra, and E.~P. Simoncelli,
\newblock ``Density modeling of images using a generalized normalization
  transformation,''
\newblock in {\em Proc. ICLR}, 2016.

\bibitem{compressai}
J. B{\'e}gaint, F. Racap{\'e}, S. Feltman, and A. Pushparaja,
\newblock ``{CompressAI}: {A} {P}y{T}orch library and evaluation platform for
  end-to-end compression research,''
\newblock {\em arXiv preprint arXiv:2011.03029}, 2020.

\bibitem{vime90k}
T. Xue, B. Chen, J. Wu, D. Wei, and W.~T. Freeman,
\newblock ``{V}ideo enhancement with task-oriented flow,''
\newblock {\em Int. J. Comput. Vis.}, vol. 127, no. 8, pp. 1106--1125, Feb.
  2019.

\bibitem{coco2014}
T.-Y. Lin, M. Maire, S. Belongie, L. Bourdev, R. Girshick, J. Hays, P. Perona,
  D. Ramanan, C.~L. Zitnick, and P. Dollár,
\newblock ``Microsoft {COCO}: {C}ommon objects in context,'' 2014.

\bibitem{kodak_dataset}
E. Kodak,
\newblock ``Kodak lossless true color image suite ({P}hoto{CD} {PCD}0992),''
  \url{http://r0k.us/graphics/kodak}.

\bibitem{theil1972statistical}
H. Theil,
\newblock {\em Statistical decomposition analysis: With applications in the
  social and administrative sciences},
\newblock North-Holland Publishing Company, 1972.

\end{thebibliography}

\end{document}